\documentclass[runningheads]{svmult}

\usepackage{makeidx}   
\usepackage{graphicx}  
\usepackage{subeqnar}  
\usepackage{multicol}  
\usepackage{physprbb}  
\makeindex             

\begin{document}
\title*{The optical afterglow and host galaxy of GRB~000926
}
\titlerunning{GRB~000926}
\author{Johan P.U. Fynbo\inst{1}
\and Javier Gorosabel\inst{2}
\and Palle M\o ller\inst{1}
\and Jens Hjorth\inst{3}
\and Michael I. Andersen\inst{4}
\and Mathias P. Egholm \inst{5}
\and Brian L. Jensen \inst{3}
\and Holger Pedersen \inst{3}
\and Bjarne Thomsen \inst{5}
\and Michael Weidinger \inst{5}
}

\authorrunning{Johan Fynbo et al.}

\institute{ESO, Garching, Germany \\
\and DSRI, Copenhagen, Denmark \\
\and Copenhagen Observatory, Denmark \\
\and Division of Astronomy, Oulu University, Finland \\
\and IFA, \AA rhus, Denmark
}    
    
\maketitle      

\begin{abstract}
In this paper we illustrate with the case of GRB~000926 how Gamma Ray
Bursts (GRBs)
can be used as cosmological lighthouses to identify and study 
star forming galaxies at
high redshifts. The optical afterglow of the burst was located with
optical imaging at the Nordic Optical Telescope 20.7 hours after the
burst. Rapid follow-up spectroscopy
allowed the determination of the redshift of the burst and
a measurement of the host galaxy HI-column density in front of the burst.
With late-time narrow band Ly$\alpha$ as well as
broad band imaging, we have studied the emission from the host galaxy
and found that it is a strong Ly$\alpha$ emitter in a state of active 
star formation. 
\end{abstract}

\section{Introduction}
Although the nature of the ``central engines'' of GRBs still are
a subject of intense debate
it is now well established that the majority of the long duration
GRBs occur in star forming galaxies at cosmological redshifts 
(see Van Paradijs, Kouveliotou \& Wijers 2000 for a review).

In this paper we focus on GRB~000926. This long duration burst was 
detected on September 26.9927 (UT) 2000 by 
three instruments (Ulysses, Konus and NEAR) in the Interplanetary 
Network (IPN, e.g. Klebsabel 
et al. 1982; Hurley et al. 2000), and localized to a
35 arcmin$^2$ error box which was circulated via the GRB Coordinates
Network (GCN)\footnote{
{\tt \small http://gcn.gsfc.nasa.gov/gcn/}}
20.3 hours after the burst (Hurley 2000).
GRB~000926 was well studied over a wide range of the
electromagnetic spectrum (Fynbo et al. 2001; Price et al. 2000, 
Piro et al. 2001; Harrison et al. 2001).

\section{Identification of the afterglow}
\begin{figure}[ht]
\begin{center}
\includegraphics[width=0.7\textwidth]{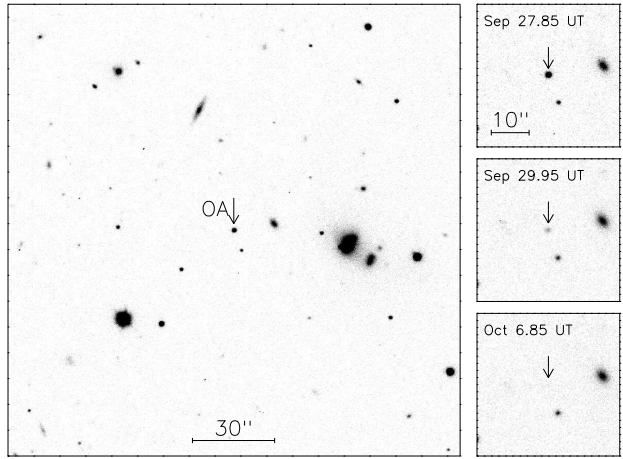}
\end{center}
\caption[]{{\it Left panel\/}: The R--band image of the OA 
taken 17 minutes after the release of the IPN error box
coordinates. The OA is marked with an arrow.
{\it Right panels\/}: Three smaller R--band images
at three epochs showing the decline of the OA.}
\label{OT}
\end{figure}

Optical follow-up observations started at the Nordic Optical Telescope
17 minutes after the release of the IPN error box coordinates (Hurley 2000).
In the main panel of Fig.~\ref{OT}, we show an R--band image of the Optical 
Afterglow (OA). The OA is marked with an arrow. The three small panels 
show the fading of the afterglow during the following days.

\section{Spectroscopy}
\begin{figure}[b]
\begin{center}
\includegraphics[width=0.35\textwidth,clip=,angle=-90]{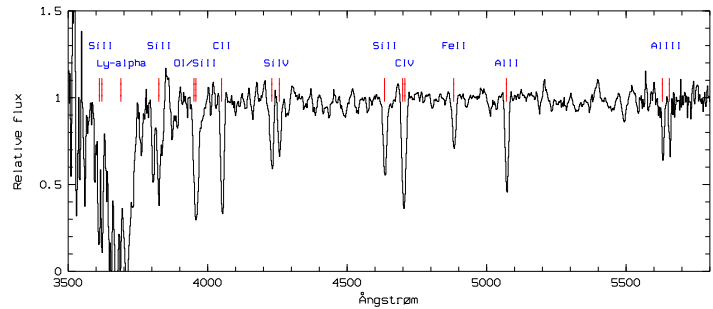}
\end{center}
\caption[]{The spectrum of GRB~000926 showing strong metal
absorption lines as well as a damped Ly$\alpha$ absorption
line at z=2.0377.}
\label{spec}
\end{figure}

Optical spectra of the afterglow were obtained
21.7 hours and 44.4 hours after the burst. The combined spectrum
is shown in Fig.~\ref{spec}. Seen are a number of strong metal 
absorption lines from which a redshift of z=2.0377 is determined. 
We detect no lines from intervening absorbers.
In the blue end of the spectrum there is a damped absorption line 
due to neutral hydrogen from which we infer a HI
column density of around 2$\times$10$^{21}$ cm$^{-2}$. The
equivalent widths of the metal absorption lines are stronger than 
for any known Damped Ly$\alpha$ Absorber at similar redshifts 
(M{\o}ller et al. in preparation).

\section{The host galaxy}

\begin{figure}[ht]
\begin{center}
\includegraphics[width=1.0\textwidth]{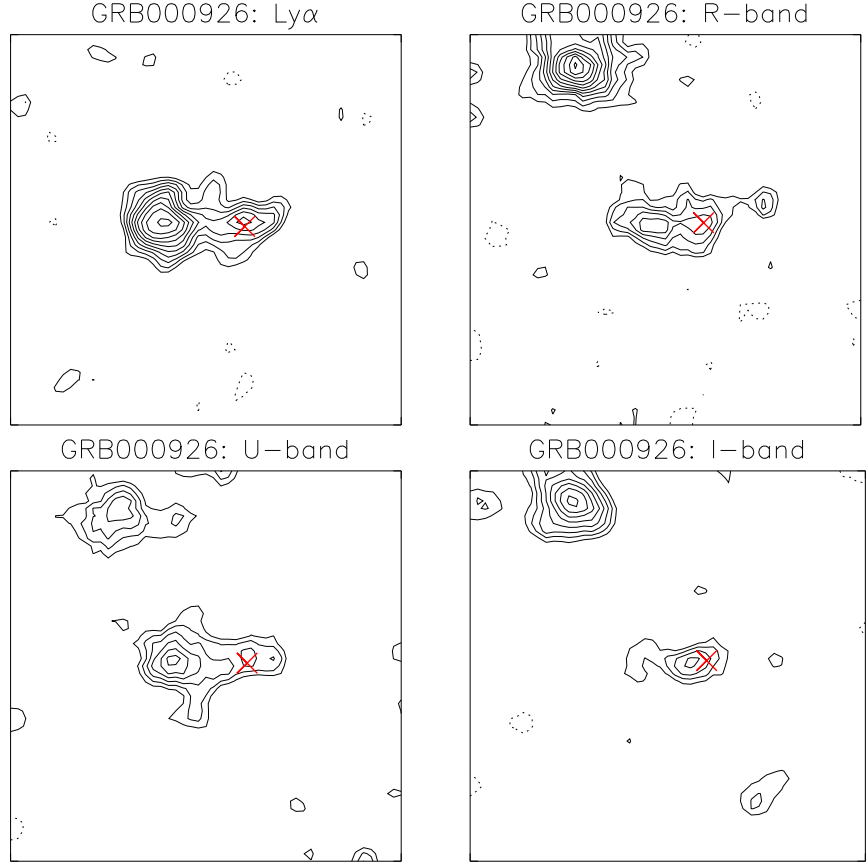}
\end{center}
\caption[]{The host galaxy of GRB~000926 as imaged in
Ly$\alpha$, U, R and I. The R-band image is taken from Fynbo et al.
(2001). East is to the left and north is up. The size of the images
is 10$\times$10 arcsec$^2$. The position of the optical afterglow is
indicated with an $\times$. The GRB occurred in the redder and fainter
western part of the Ly$\alpha$ emitting region.
}
\label{host}
\end{figure}

With deep R--band imaging obtained 1 month after the burst we
detect the host galaxy as an extended object consisting of several
compact knots. In order to study the host galaxy further, we 
obtained a special narrow (fwhm 45\AA) band filter designed to cover 
Ly$\alpha$ at the GRB redshift. In May 2000 we obtained 12 hours of
imaging in the narrow filter as well as 7 hours of imaging in the
U filter and 3 hours of imaging in the I filter at the Nordic
Optical Telescope. We detect the host galaxy in all bands
(Fig.~\ref{host}). The host galaxy is a relatively strong Ly$\alpha$
emitter (we detect one brighter Ly$\alpha$ emitter in the field).
About 65\% of the Ly$\alpha$ emission comes from the eastern knot
of the host and the remaining 35\% from the western knot in which the 
GRB occured. Although the signal-to-noise ratio is low, the western 
knot seems to be brightest in the I-band, and it must hence be redder than 
the other components. This indicates either more extinction or the 
presence of an older stellar population.

\section{Summary}
GRB~000926 occurred in a star forming galaxy at a redshift of z=2.0377.
The optical spectrum of the afterglow shows strong metal absorption
as well as a damped Ly$\alpha$ absorption from metal enriched gas in 
the host galaxy of the burst.
The galaxy is a strong Ly$\alpha$ emitter consisting of at least two 
compact knots. The GRB occurred in the western and reddest of the knots.

\section*{Acknowledgments}
The data presented here have been taken using ALFOSC, which is owned
by the Instituto de Astrofisica de Andalucia (IAA) and operated at the
Nordic Optical Telescope under agreement between IAA and the NBIfAFG
of the Astronomical Observatory of Copenhagen.
Nordic Optical Telescope is operated on the island of La Palma 
jointly by Denmark, Finland, Iceland, Norway, and Sweden, in the
Spanish Observatorio del Roque de los Muchachos of the 
Instituto de Astrofisica de Canarias. MPE and MW acknowledge
support from the ESO Directors Discretionary Fund. JG
acknowledges the receipt   of a Marie  Curie Research Grant
from the European Commission. This work was
supported by the Danish Natural Science Research Council (SNF).


\begin{thebibliography}{8.}
\addcontentsline{toc}{section}{References}

\bibitem{Fynbo}J.U. Fynbo, J. Gorosabel., T.H. Dall, et al.: A\&A
\textbf{373}, 796 (2001)
\bibitem{Harrison}F.A. Harrison, S.A. Yost, R. Sari, et al.: ApJ
\textbf{559}, 123 (2001)
\bibitem{Hurley2}K. Hurley, 2000, GCN 801 and 802
\bibitem{Hurley} K. Hurley, T. Cline, E. Mazets, et al.: ApJL
\textbf{534}, L23 (2000)
\bibitem{Kleb} R. Klebsabel, W. Evans, D. Laros, et al.: ApJL \textbf{259}, L51 (1982)
\bibitem{vP} J. van  Paradijs, C. Kouveliotou, R.A.M.J. Wijers: ARA\&A
\textbf{38}, 379 (2000)
\bibitem{Piro}L. Piro, G. Garmire, M.R.  Garcia, et al.: ApJ
\textbf{558}, 442 (2001)
\bibitem{Price}P.A. Price, F.A. Harrison, T.J. Galama, et al.: ApJL
\textbf{549}, 7 (2000)

\end{thebibliography}
\end{document}